\documentclass[a4paper,12pt]{article}

\title{\LARGE\bf Dipolar SLEs.}
\date{}
\author{}

\newcommand{\vev}[1]{\langle #1 \rangle}
\newcommand{\ket}[1]{| #1 \rangle}
\newcommand{\bra}[1]{\langle #1 |}

\def\debut{\begin{eqnarray}}
\def\fin{\end{eqnarray}}

\usepackage[final]{graphicx}

\usepackage{amssymb} 
\usepackage{oldgerm} 

\begin{document}
\maketitle

\vspace{-1.5 truecm}

\centerline{\large M. Bauer\footnote{Email: 
      michel.bauer@cea.fr}, D.
Bernard\footnote{Member of the CNRS; email:
  dbernard@spht.saclay.cea.fr} and J. Houdayer\footnote{Member of the
  CNRS; email: jerome.houdayer@polytechnique.org}}

\vspace{.3cm}

\centerline{\large Service de Physique Th\'eorique de Saclay}
\centerline{CEA/DSM/SPhT, Unit\'e de recherche associ\'ee au 
CNRS\footnote{URA 2306 du CNRS}}
\centerline{CEA-Saclay, 91191 Gif-sur-Yvette, France}


\vspace{1.0 cm}

\begin{abstract}
  We present basic properties of {\it Dipolar SLEs}, a new version of
  stochastic Loewner evolutions (SLE) in which the critical interfaces
  end randomly on an interval of the boundary of a planar domain. We
  present a general argument explaining why correlation functions of
  models of statistical mechanics are expected to be martingales and
  we give a relation between dipolar SLEs and CFTs.  We compute SLE
  excursion and/or visiting probabilities, including the probability
  for a point to be on the left/right of the SLE trace or that to be
  inside the SLE hull. These functions, which turn out to be harmonic,
  have a simple CFT interpretation. We also present numerical
  simulations of the ferromagnetic Ising interface that confirm both
  the probabilistic approach and the CFT mapping.

\end{abstract}


\vskip 1.5 truecm


\bigskip

Two types of stochastic Loewner evolutions (SLE) have been thoroughly
studied, the chordal and the radial SLEs \cite{Schramm0, RohdeSchramm,
  LSW, Lawler}. The former describe random curves joining two points
on the boundary of a simply connected planar domain while the latter
describe curves joining one point on the boundary to a point in the
bulk of the domain. They correspond to two inequivalent normalizations
of conformal maps between simply connected domains in $\mathbb{C}$.
Using geometrical constraints, we realized in ref.\cite{zigzag} that
there was a third inequivalent normalization and thus that there was
yet another process, which we called the {\it dipolar SLE}, with all
properties required to define an SLE, depending as usual on a real
positive parameter $\kappa$.  Concretely, dipolar SLEs describe random
curves in a simply connected planar domain which start at a point
$x_0$ on the boundary and are stopped the first time they hit the
boundary on a specified interval $[x_+,x_-]$ not containing $x_0$.
They generalize chordal SLEs. One of the virtues of dipolar versus
chordal SLEs is that the hull does not fill the full domain for
$\kappa > 4$. This makes the dipolar geometry physically appealing.
It is for instance the most natural one to describe several interface
properties already computed, such as Cardy's formula for percolation
\cite{Cardy,Smirnov}.  The chordal case corresponds to the limit when
$x_+$ and $x_-$ merge together, and in this limit the hull
invades the full domain for $\kappa>4$.

The aim of the following is to introduce dipolar SLEs and to present
their basic properties. 

Section \ref{sec:dipolarSLEs} gives the precise definition of dipolar
SLEs.
  
Section \ref{sec:statmechmart} is a digression of general nature which
emphasizes why (conditional) correlation functions of models of
statistical mechanics are martingales for appropriate stochastic
processes. This can be used to elucidate the link between SLEs and
conformal field theories (CFTs).
  
Section \ref{sec:cftconn} applies these general ideas to build the CFT
interpretation of dipolar SLEs, identifying in particular the boundary
operators acting at $x_0$ and $x_\pm$.
  
Section \ref{sec:bulkvisprob} is devoted to the computation of some
basic dipolar bulk probabilities for $\kappa >4$, namely the
probability for a point to be on the left of (resp. on the right of)
or swallowed by the SLE hull. These probabilities are computable
because they turn out to be harmonic solutions of the
general martingale equation~(\ref{martineq}). Why this equation has
interesting harmonic solutions remains to be explained. In fact, the
probabilities for a point to be swallowed by the hull from the left or
from the right for $\kappa >4$ are non-harmonic, and so are the
probabilities to be on the left or on the right of the SLE hull for
$\kappa < 4$. The explicit computation of these probabilities has
eluded us.
  
Section \ref{sec:limcase} is devoted to the limiting case $\kappa=4$
and its relation to free field theory with an alternation of
appropriate boundary conditions, namely Dirichlet between $x_-$ and
$x_0$, and between $x_0$ and $x_+$ (but with two different boundary
values), and Neumann between $x_+$ and $x_-$.
  
In Section \ref{sec:boundexc} we compute, for arbitrary $\kappa$, the
distribution of the hitting point of the hull on the interval $[x_-,x_+]$
that does not contain $x_0$. We also give a detailed CFT derivation of
the result.

In Section \ref{sec:simul} we present Monte Carlo simulations of the
Ising ferromagnet with specific boundary conditions corresponding to
the case $\kappa=3$. The distribution of the interface endpoint
agrees very well with the theoretical prediction presented in this
article. This confirms the validity of the mapping to the Ising model.

\bigskip

\section{Dipolar SLEs.}  \label{sec:dipolarSLEs}

As for any SLEs, the random curves, often called SLE traces, are
encoded into a family of conformal maps $g_t(z)$ parameterized by a
`time' $t$. When the traces are simple curves, the map $g_t$
uniformizes the complement of the portion of the curve
$\gamma_{[0,t]}$ in the domain on which the curve is growing back into
this domain. Such maps exist by the Riemann mapping theorem and they
are defined up to $SL_2(\mathbb{R})$ transformations by global
conformal symmetry.  Thus we need three conditions to fully specify
them. For dipolar SLEs, we choose two points on the boundary, $x_\pm$,
and impose that $g_t(x_\pm)=x_\pm$ and $g_t'(x_+)=g_t'(x_-)$ for each
$t$. The process is then defined by specifying the stochastic
evolution of $g_t$ via a Loewner equation.  The dipolar SLE maps
describe curves starting at a point $x_0\neq x_\pm$ on the boundary
and ending on the boundary at a random point on the interval between
the two fixed points that does not contain $x_0$. The crucial point is
that for a simply connected domain, the group of conformal
automorphisms fixing two boundary points is a one parameter Lie group
isomorphic to the additive group of real numbers. Hence, there is a
canonical definition of a Brownian motion between $x_-$ and $x_+$
starting at $x_0$, at least when the corresponding boundary interval
is a simple curve. See \cite{Schramm0, RohdeSchramm, LSW, Lawler} for
basic -- and not so basic -- material on SLEs.

To be more concrete, 
let $\mathbb{S}_\Delta=\{z\in \mathbb{C},\ 0<\Im{\rm m}\, z<\pi
\Delta\}$ be the strip of width $\pi\Delta$ which is the geometry
adapted to dipolar SLEs.
By definition, the dipolar Loewner equation in $\mathbb{S}_\Delta$
reads:
\begin{eqnarray}
\partial_t g_t(z)=\frac{1/\Delta}{\tanh\left({(g_t(z)-\xi_t)/2\Delta}\right)},
\quad g_{t=0}(z)=z,
\label{Sdipol}
\end{eqnarray}
with $\xi_t=\sqrt{\kappa}\, B_t$ with $B_t$ a normalized Brownian
motion and $\kappa$ a real positive parameter so that ${\bf
  E}[\xi_t\,\xi_s]=\kappa\,{\rm min}(t,s)$.  The two boundary fixed
points are $x_\pm=\pm\infty$ and the starting point $x_0$ is the
origin.  The maps $g_t$ are normalized to fix $x_\pm$.  For fixed $z$,
$g_t(z)$ is well-defined up to the time $\tau_z$ for which
$g_{\tau_z}(z)=\xi_{\tau_z}$. Times $\tau_z$ are called swallowing
times.  In the limit $\Delta\to \infty$ we recover the chordal SLEs.
$\Delta$ is simply a dilatation factor. Unless otherwise specified, we
will set $\Delta=1$ in the following and we will look at
dipolar SLEs in the strip $\mathbb{S}_1$.

As for chordal or radial SLEs \cite{Schramm0,RohdeSchramm,LSW,
  Lawler}, the SLE hull is reconstructed from $g_t$ by
$\mathbb{K}_t=\{z\ \mathrm{such~that}\ \tau_z\leq t\}$ and the SLE trace
$\gamma_{[0,t]}$ by
$\gamma(t)=\lim_{\epsilon\to0^+}g_t^{-1}(\xi_t+i\epsilon)$.  It is
known that $\gamma_{[0,t]}$ is almost surely a curve.  It is non-self
intersecting and it coincides with $\mathbb{K}_t$ for $0<\kappa\leq
4$, while for $4<\kappa<8$ it possesses double-points and it does not
coincide with $\mathbb{K}_t$ which is then the set of points swallowed
up to time $t$. For $\kappa \geq 8$, the trace is space filling.
However, in contrast to the chordal SLE case, for dipolar SLEs, the hull
does not fill the entire domain even for $\kappa \geq 8$.

It will be convenient to consider the map which maps the tip of the
SLE trace back to its starting point. Thus we translate $g_t$ and let
$f_t(z)=g_t(z)-\xi_t$ so that $f_t(\gamma(t))=0$.  The dipolar
stochastic equation is simply:
$$
df_t(z)= \Delta^{-1}\, \coth (f_t(z)/2\Delta) + d\xi_t.
$$
The maps $f_t$ are such that, for $s>t$, $f_s\circ f_t^{-1}$ is
distributed as $f_{s-t}$.

Eq.(\ref{Sdipol}) may be integrated in the simple deterministic case
with $\xi_t$ constant. Then $\cosh(g_t(z)-\xi)/2=
e^{t/2}\,\cosh(z-\xi)/2$ for $\Delta=1$.  The trace $\gamma(t)$ is
determined by $\cosh(\gamma(t)-\xi)/2= e^{-t/2}$, so that it starts on
the real axis at $\xi$ and stops when it touches the upper boundary of
the strip at the point $i\pi+\xi$. As expected, all points at the left
of the trace are mapped to the fixed point at $-\infty$ by $f_t$ as
$t\to\infty$, while those on the right of the curve are mapped into
the other fixed point $+\infty$.

\begin{figure}[htbp]
\begin{center}
\includegraphics[width=0.8\textwidth]{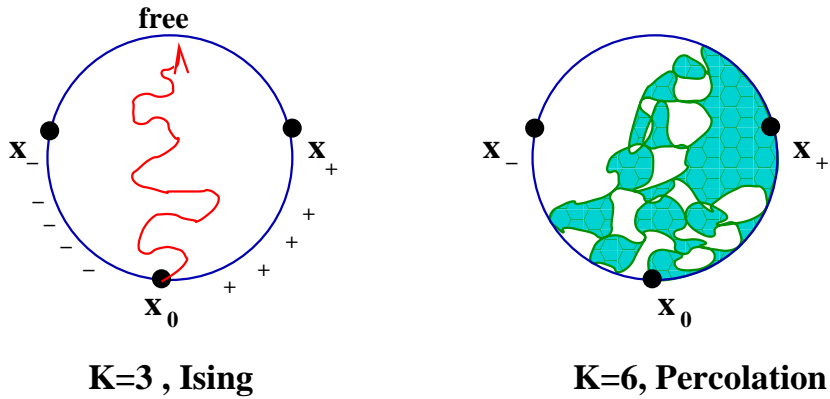}
\caption{Two schematic examples of dipolar SLEs.}
\label{fig:exdip}
\end{center}
\end{figure}

Let us return to the case when $\xi_t$ is a Brownian motion.  A simple
probabilistic argument, explained in \cite{zigzag}, shows that the
probabilities for the trace to touch the upper boundary in finite time
vanishes. So, the trace only touches this boundary at infinite
time. This amounts to stop the process when the trace hits the upper
boundary, a criterion which is invariant under time
reparametrizations of the evolution. Since locally the dipolar SLE trace
looks the same as the chordal or radial SLE traces
\cite{RohdeSchramm}, this leads to the following picture. For
$\kappa\leq 4$, the hull coincides with the trace. The latter does not
touch the lower boundary, ie. the real axis $\mathbb{R}$, but stops
when it hits the upper boundary, ie.  $i\pi+\mathbb{R}$, at same
random point. For $\kappa>4$, the trace is not a simple curve and
points in $\mathbb{S}_1$ are swallowed, the set of which form the SLE
hull.  The hull then intersects the lower boundary, ie. the real axis,
and the trace hits this boundary an infinite number of times but hits
only once the upper boundary, again at some random point, and stops
there.  See Figure (\ref{fig:exdip}).

\section{Statistical mechanics and martingales.}\label{sec:statmechmart}

This section is a recreative interlude in which we explain why
(conditional) correlation functions of models of statistical mechanics
are martingales for appropriate stochastic processes. As a very
general statement this remark may look tautological but it is
nevertheless quite instructive. In particular it provides a key to
decipher the relation between SLEs and CFTs.

Let ${\cal C}$ be the configuration space of a statistical model. For
simplicity we assume ${\cal C}$ to be discrete and finite
but as large as desired. Let $W_c$ be the Boltzmann weights and 
$Z$ the partition function, $Z=\sum_{c\in{\cal C}}\ W_c$.

We imagine having introduced a family ${\cal Q}_t$ of partitions of the
configuration space whose elements ${\cal C}_{\alpha_t}$ 
are labeled by indices $\alpha_t$:
$$ {\cal C}= \bigcup_{\alpha_t}{\cal C}_{\alpha_t},
\qquad {\cal C}_{\alpha_t}\ {\rm disjoints}.
$$
The index $t$, which will be identified with `time', labels the
partitions. By convention ${\cal Q}_0$ is the trivial partition with
${\cal C}$ as its single piece.
We assume these partitions to be finer as $t$ increases,
which means that for any $s>t$ and any element ${\cal C}_{\alpha_t}$
of the partition at time $t$ there exist elements of ${\cal Q}_s$
which form a partition of ${\cal C}_{\alpha_t}$. An example of such
partitions in case of spin statistical models consists in
specifying the values of local spin variables at an increasing number of
lattice points. Block spin clustering used in renormalization group is 
another way to produce such partitions.
A SLE inspired example consists in specifying the
shapes and the positions of interfaces of increasing lengths.

We define the restricted partition function $Z_{\alpha_t}$ by
$$
Z_{\alpha_t} \equiv \sum_{c\in{\cal C}_{\alpha_t}} W_c~.
$$
Since restricting the summation to a subset amounts to impose some
condition on the statistical configurations, $Z_{\alpha_t}$ is the
partition function conditioned by the knowledge specified by 
${\cal C}_{\alpha_t}$.

To define a stochastic process we have to specify the probability
space and a filtration on it. The simplest choice is ${\cal C}$ as a
probability space equipped with its canonical $\sigma$-algebra, ie. the 
one generated by all its singletons, and with the probability measure
induced by the Boltzmann weights, ie. ${\bf P}[\{c\}]= W_c/Z$.
In particular the probability of the event ${\cal C}_{\alpha_t}$ is
the ratio of the partition functions
$$ {\bf P}[{\cal C}_{\alpha_t}]= Z_{\alpha_t}/Z.$$
To any partition ${\cal Q}_t$ is associated a $\sigma$-algebra on
${\cal C}$, ie. the one generated by the elements of this partition.
Since these partitions are finer as `time' $t$ increases, it induces a
filtration ${\cal F}_t$ on ${\cal C}$ equipped with its probability
measure. 

Now, given an observable ${\cal O}$ of the statistical model, 
ie. a function $c\to {\cal O}_c$ on the configuration space, 
we can define its conditional average
$$ \vev{ {\cal O} }_t \equiv {\bf E} [ {\cal O}|{\cal F}_t].$$
By construction, $\vev{ {\cal O} }_t$ is a function which is constant
on any element ${\cal C}_{\alpha_t}$ of the partition ${\cal Q}_t$ and 
takes values
$$ \vev{ {\cal O} }_t \vert_{ {\cal C}_{\alpha_t} }  =
\frac{1}{Z_{\alpha_t}}
\sum_{c\in{\cal C}_{\alpha_t}} {\cal O}_c\ W_c. $$
This is simply the statistical average conditioned on the knowledge
specified by ${\cal C}_{\alpha_t}$.

By construction, $ \vev{{\cal O}}_t$ is a (closed) martingale with
respect to ${\cal F}_t$. Indeed, for $t>s$,
$$
 {\bf E}[\vev{ {\cal O} }_t |{\cal F}_s]=
 {\bf E} [ {\bf E}[ {\cal O}|{\cal F}_t] |{\cal F}_s]
={\bf E}[ {\cal O}|{\cal F}_s]=\vev{{\cal O}}_s,
$$
where we used standard properties of conditional expectations and
that ${\cal F}_t\subset {\cal F}_s$ for $t>s$.
In particular, one may verify that its average is time independent and
equals to the statistical average:
\begin{eqnarray}
{\bf E}[\vev{ {\cal O} }_t]
&=&\sum_{\alpha_t} {\bf P}[ {\cal C}_{\alpha_t}]\ 
\vev{ {\cal O} }_t\vert_{ {\cal C}_{\alpha_t} } \nonumber \\
&=& \frac{1}{Z} \sum_{c\in{\cal C}} {\cal O}_c\ W_c = \vev{{\cal O}}.
\nonumber
\end{eqnarray}

This observation formally applies to critical interfaces and hence to
SLEs. The remarkable observation made by O. Schramm 
is that conformal invariance implies that the
filtration associated to the partial knowledge of the interface is
that of a continuous martingale, i.e. that of a Brownian motion if
time is chosen cleverly. The only parameter is $\kappa$. The physical
parameters of the CFT, for instance the central charge, can be
retrieved by imposing that the correlation functions $\vev{ {\cal O}
}_t$ conditioned by the knowledge of ${\cal F}_t $ be martingales.
The CFT situation is particularly favorable in that going from $\vev{
  {\cal O} }_t$ to $\vev{ {\cal O} }$ is pure kinematics.

\section{CFT connections.}\label{sec:cftconn}

Connection with conformal field theories can be done using the CFT
operator formalism. The latter is simpler if the boundary CFT is
considered in the upper half plane
$\mathbb{H}=\{z \in \mathbb{C}, \Im {\rm m}\, z > 0\}$. 

SLE processes are transported onto any simply connected domain by
conformal transformations, by definition.  In $\mathbb{H}$, we fix the
boundary points $x_\pm$ to be $\pm1$ and $x_0$ to be the origin.  The
conformal map uniformizing $\mathbb{S}_\Delta$ onto $\mathbb{H}$ is
$\varphi(w)=\tanh(w/2\Delta)$. Let $\hat g_t$ be the dipolar SLE map
in $\mathbb{H}$, $\hat g_t\circ \varphi = \varphi\circ g_t$.  The
stochastic Loewner equation for dipolar SLE in
$\mathbb{H}$ reads \cite{zigzag}:
$$
\partial_t \hat g_t(z)=\left(\frac{1-\hat g_t(z)^2}{2}\right)\,
\left(\frac{1-\hat g_t(z)\tanh \xi_t/2}{\hat g_t(z)-\tanh \xi_t/2}\right),
\quad \hat g_{t=0}(z)=z. 
$$
The maps $\hat g_t$ are normalized to fix $x_\pm=\pm 1$  and to have equal
derivatives at these two points : $\hat g'_t(\pm 1)=e^{-t}$. 

The map $f_t$ can also be transported in the upper half plane to
produce a map $\hat f_t=\tanh(f_t/2\Delta)$, which also fixes
$x_\pm=\pm 1$ but which maps the tip of the SLE trace in $\mathbb{H}$
to the origin. Its expression is:
$$ \hat f_t(z) = 
\frac{\hat g_t(z)-\tanh\xi_t/2}{1-\hat g_t(z) \tanh\xi_t/2}.
$$
The stochastic differential equation that $\hat f_t$ satisfies
directly follows from that obeyed by $f_t$.

As for the chordal and radial cases \cite{BaBe}, the connection
between SLEs and CFTs may be established by associating to $\hat
f_t$ an operator $\hat G_t$ which implements this conformal
transformation in the CFT Hilbert space. Since $\hat f_t$ fixes
the point $x_+$, or $x_-$, $\hat G_t$ may be constructed as an element
of the enveloping algebra of the appropriate Borel sub-algebra of the
Virasoro algebra. See \cite{BaBe} for details. It intertwines between
primary fields, say $\Phi_{h,\bar h}(z,\bar z)$ of conformal dimensions
$(h,\bar h)$, and their images under $\hat f_t$. Namely,
$$
\hat G_t^{-1}\, \Phi_{h,\bar h}(z,\bar z)\, \hat G_t
= [\hat f_t'(z)]^h [\overline{\hat f_t'(z)}]^{\bar h}\
 \Phi_{h,\bar h}(\hat f_t(z),\overline{\hat f_t(z)}).
$$
The stochastic differential equation that $\hat G_t$ satisfies directly
follows from that of $\hat f_t$, or $\hat g_t$. It reads \cite{zigzag}:
\begin{eqnarray}
\hat G_t^{-1}\, d\hat G_t = (-2 W_{-2} +\frac{\kappa}{2}W_{-1}^2)dt -
W_{-1}d\xi_t,  \label{Gstoc}
\end{eqnarray}
where $W_{-2}$ and $W_{-1}$ are elements of the Virasoro 
algebra\footnote{We denote by $L_n$ the generators of the Virasoro algebra
  with commutation relations
  $[L_n,L_m]=(n-m)L_{m+n}+\frac{c}{12}n(n^2-1)\delta_{n+m,0}$ and $c$
  is the central charge.}, 
$$
W_{-2} = \frac{1}{4}(L_{-2}-L_0),\quad
W_{-1}=\frac{1}{2}(L_{-1}-L_1).
$$
In this algebraic setting, the differences between radial and dipolar
SLEs may be viewed as coming from a different choice of real forms in the
Virasoro algebra.

As in the chordal and radial SLEs, the key point is now the
construction of a generating function of local martingales which is
obtained using a representation of the Virasoro algebra degenerate at
level two. 

Let $\ket{\omega}$ be the highest weight vector of the irreducible
Virasoro module with central charge $c$ and conformal weight $h_{1;2}$,
$$ c= \frac{(\kappa-6)(8-3\kappa)}{2\kappa},\quad
h_{1;2}=\frac{6-\kappa}{2\kappa},$$
then 
\begin{eqnarray}
M_t \equiv e^{+2h_{0;1/2}\, t}\, \hat G_t\ket{\omega}
\label{martin}
\end{eqnarray}
is a local martingale, with
$$ h_{0;1/2}= \frac{(6-\kappa)(\kappa-2)}{16\kappa}.$$

This follows from the null vector relation
$(-2L_{-2}+\frac{\kappa}{2}L_{-1}^2)\ket{\omega}=0$. 
Indeed, a simple rearrangement leads to
$$-2W_{-2}+\frac{\kappa}{2}W_{-1}^2=\frac{1}{4}(-2L_{-2}+\frac{\kappa}{2}L_{-1}^2)
+\frac{2-\kappa}{4}L_0 +\frac{\kappa}{8}(L_1^2-2L_{-1}L_1),$$
so that
$(-2W_{-2}+\frac{\kappa}{2}W_{-1}^2)\ket{\omega}=-2h_{0;1/2}\ket{\omega}$
and 
$$\hat G^{-1}d\hat G_t\ket{\omega}= -dt\,2h_{0;1/2}\ket{\omega}
 + d\xi_t\, W_{-1}\ket{\omega}.$$

In particular, by projecting this local martingale on vectors $\bra{v}$
and assuming appropriate boundedness conditions, we get that the
expectations ${\bf E}[e^{2h_{0;1/2}t}\, \bra{v}\hat G_t \ket{\omega}]$
are time independent.

The prefactor $e^{+2h_{0;1/2}t}$ accounts for the insertion of two
boundary conformal fields, each of dimension $h_{0;1/2}$,
localized at the two fixed points $x_\pm$ since $\hat f_t'(x_+)\hat
f_t'(x_-)=e^{-2t}$. Alternatively, the local martingales can be written
as the ratio of two correlation functions with insertions of the
boundary operator $\psi_{1;2}$ creating the state $\ket{\omega}$ at
$x_0$ and of the operators $\psi_{0;1/2}$ at the fixed points $x_\pm$:
\begin{eqnarray}
\frac{\vev{\psi_{0;1/2}(x_-)\psi_{0;1/2}(x_+)\, 
{\cal O}\, \hat G_t\, \psi_{1;2}(x_0)}}{
\vev{\psi_{0;1/2}(x_-)\psi_{0;1/2}(x_+)\, 
\hat G_t\, \psi_{1;2}(x_0)}}
\label{OvevG}
\end{eqnarray}
for any operator ${\cal O}$.  Since martingales are key ingredients
for computing probabilities, this statement implies that dipolar SLE
probabilities will be expressible in terms of CFT correlation
functions with insertions of these boundary operators.

By conformal invariance, this last expression can be transported to
any simply connected planar domain $\mathbb{S}$. If $g_t$ is the
conformal map describing the growth of the dipolar SLE hulls in
$\mathbb{S}$, mapping the tip of the SLE trace to
$x_0\in\partial\mathbb{S}$, then the ratios
\begin{eqnarray}
\frac{\vev{\psi_{0;1/2}^{g_t}(x_-)\psi_{0;1/2}^{g_t}(x_+)\, 
{\cal O}^{g_t}\,\psi_{1;2}(x_0)}}{
\vev{\psi_{0;1/2}^{g_t}(x_-)\psi_{0;1/2}^{g_t}(x_+)\psi_{1;2}(x_0)}}
\label{OvevGbis}
\end{eqnarray}
are local martingales. Here $\psi^{g_t}_{0;1/2}$, $\psi^{g_t}_{1;2}$
or ${\cal O}^{g_t}$ denote the images of the corresponding fields by
$g_t$, ie. their pull-back by $g_t$. As explained in the previous
section, this statement has a natural explanation in basic statistical
mechanics. Although this construction can be seen merely as a trick to
construct martingales, the previous section explains why the
martingale property is what ensures that the CFT identified in this
way is precisely the critical continuum limit of the statistical
mechanics model that produces the interface described by SLE.  
\medskip

{\it --- Examples:}

Dipolar SLEs has a simple statistical interpretation for $\kappa=3$
which corresponds to the Ising model with $c=1/2$.  Then $\psi_{1;2}$
with $h_{1;2}=1/2$ is the boundary condition changing operator between
spin $(+)$ and spin $(-)$ fixed boundary conditions, while
$\psi_{0;1/2}$ with $h_{0;1/2}=1/16$ is the boundary condition
changing operator between fixed and free boundary conditions
\cite{cardybcc}.  Hence, along the boundary of the domain one
encounters the boundary conditions fixed $(+)$, then free, and then
again fixed but $(-)$, as depicted in Fig.(\ref{fig:exdip}). It is
thus clear that we need one operator $\psi_{1;2}$ and two operators
$\psi_{0;1/2}$ to describe this system.

As we shall see in the following sections, 
the case $\kappa=4$ which corresponds to a free bosonic field 
with $c=1$ also has a simple interpretation with an alternation of
Dirichlet and Neumann boundary conditions. 

For the 3-state Potts model we may propose the following
interpretation. The microscopic spin variables take three possible
values $(1)$, $(2)$ and $(3)$ related by $S_3$ symmetry. The 3-state
Potts model corresponds to $\kappa=10/3$ with central charge $c=4/5$.
Then $h_{1;2}=2/5$ and $h_{0;1/2}=1/15$.  The operator $\psi_{1;2}$ is
the boundary operator between a fixed boundary condition with all
spins $(1)$ and a mixed boundary condition $(2+3)$ with a mixture
of spins $(2)$ and $(3)$. The operator $\psi_{0;1/2}$ is the boundary
operator between the boundary conditions $(1)$ and $(1+2)$. It is also
the lowest boundary primary operator generated by a change
of mixed boundary conditions from $(1+2)$ to $(2+3)$, see
ref.\cite{cardybcc}.  Thus the dipolar SLE with $\kappa=10/3$ should
correspond to the 3-Potts models with the succession of boundary
conditions, fixed $(1)$, mixed $(2+3)$ and mixed $(1+2)$.

Unitary minimal CFTs with $c=1-6/m(m+1)$ with $m$ integer correspond
to two values of $\kappa$ related by duality: $\kappa=4(m+1)/m$ or
$\kappa=4m/(m+1)$.  Since the identification of $\psi_{0;1/2}$ depend
on the parity of $m$ --- ie. $h_{0;1/2}=h_{m/2;m/2}$ for $m$ even,
$\kappa\geq 4$, but $h_{0;1/2}=h_{(m+1)/2;(m+1)/2}$ for $m$ odd,
$\kappa\leq 4$, --- a simple microscopic interpretation is possible
only for one of the two choices of $\kappa$. For instance, although
$\kappa=16/3$ corresponds to the Ising model, the role of the Virasoro
representation with weight $h_{0;1/2}=5/192$ in the Ising model has
not yet been clearly identified.

\section{Bulk visiting probabilities and harmonic functions
  ($\kappa>4$).}\label{sec:bulkvisprob}

Let us now look at SLE bulk properties. We assume $\kappa>4$ and we deal
with dipolar SLEs in the strip $\mathbb{S}_1$.
We shall evaluate the following probabilities:

(i) The probabilities $P_l(z,\bar z)$ (resp. $P_r(z,\bar z)$) for a
bulk point $z$ not to be swallowed by the SLE trace and to be on the
left (resp. the right) of the trace. This is the probability for the
point $z$ to be on the left (resp. the right) of the exterior frontier
of the SLE hull viewed from the boundary point $x_-$ (resp. $x_+$). It is
therefore the probability for the existence of a path joining $x_-$
(resp. $x_+$) to the boundary interval $[x_+,x_-]$ leaving the point $z$
on its right (resp.  left) and included into one cluster of the underlying
model of statistical mechanics. For $\kappa=6$ it bears some
similarities with the probability computed by Smirnov \cite{Smirnov}
to prove the equivalence between critical percolation and $SLE_6$.

(ii) The probability $P_{in}(z,\bar z)$ for the point $z$ to be
in the SLE hull. We do not distinguish the events in which the point 
has been swallowed from the right or from the left. Since this
probability is also that for the point $z$ to be in left or right
frontiers of the hull it gives informations on the shape of the hull.
These probabilities turn out to be harmonic functions for all values
of $\kappa>4$ and are proportional to the CFT correlation functions
\begin{eqnarray}
\vev{\psi_{0;1/2}(x_-)\psi_{0;1/2}(x_+)\Phi_0(z,\bar
  z)\psi_{1;2}(x_0)}
\label{btob}
\end{eqnarray}
involving a weight zero bulk primary field $\Phi_0$. 

As usual, a way to compute these probabilities is to notice that the
process $t\to P(f_t(z),\overline{f_t(z)})$ is a local martingale. 
Indeed, since $f_s\circ f_t^{-1}$, $s>t$, is independent of $f_t$ and
 distributed as $f_{s-t}$, the function $P(f_t(z),
\overline{f_t(z)})$ is the probability of the events (i), or (ii), conditioned
on the process up to time $t$ and, as such, it is a martingale.
As a consequence, the drift term in the It\^o derivative of
$P(f_t(z),\overline{f_t(z)})$ vanishes which implies that $P(z,\bar z)$
satisfies the following differential equation:
\begin{eqnarray}
\kappa \partial_z\bar\partial_{\bar z} P +
(\coth\frac{z}{2} +\frac{\kappa}{2}\partial_z)\partial_z P
+ (\coth\frac{\bar z}{2} +\frac{\kappa}{2}\partial_{\bar z})
\partial_{\bar z} P =0.
\label{martineq}
\end{eqnarray}
The main observation in this section is that, quite remarkably,
eq.(\ref{martineq}) has interesting harmonic solutions, in fact enough
harmonic solutions to compute $P_l$, $P_r$ and $P_{in}$.

The boundary conditions depend on which probabilities we are computing:\\
(i) For the probability to be on the left of the hull, this requires:
\begin{eqnarray}
 P_l(-\infty)=1,\quad P_l(+\infty)=0,\quad P_l(0)=0.
\label{bdry1}
\end{eqnarray}
Similar conditions hold for $P_r(z,\bar z)$.\\
(ii) For the probability to be in the hull, it requires:
\begin{eqnarray}
 P_{in}(-\infty)=P_{in}(+\infty)=0,\quad P_{in}(0)=1.
\label{bdry2}
\end{eqnarray}

These boundary conditions follow by noticing that if point $z$ is
swallowed at time $\tau_z$ then $\lim_{t\nearrow\tau_z}f_t(z)=0$, if
it is not swallowed but is on the left of the trace then $\lim_{t
  \nearrow\infty}f_t(z)=-\infty$, and if it is not swallowed but is on
the right of the trace then $\lim_{t \nearrow\infty}f_t(z)=+\infty$.
These conditions are such that at the stopping $\widehat \tau_z={\rm
  min}(\tau_z,\infty)$ the martingale $P(f_t(z),\overline{f_t(z)})$ projects
on the appropriate events (i) or (ii), ie.  
$ P(f_{\widehat \tau_z}(z),\overline{f_{\widehat
  \tau_z}(z)})={\bf 1}_{{\rm events}}$.  As a consequence, the
probability of these events are:
$$ {\bf P}[{\rm events}]\equiv{\bf E}[{\bf 1}_{{\rm events}}]
= {\bf E}[ P(f_{\widehat \tau_z}(z),\bar f_{\widehat \tau_z}(z))]
= P(z,\bar z).$$
The martingale property has been used in the last equality.
\medskip

{\it --- Case (i): probability to be on the left of the trace.}

The solution of the martingale equation (\ref{martineq}) satisfying the
boundary conditions (\ref{bdry1}) is given by the harmonic function
\begin{eqnarray}
P_l(z,\bar z)= 1 - \frac{\Im{\rm m}\, F(z)}{\Im{\rm m}\, F(\infty)}
 \label{probleft}
\end{eqnarray}
with 
$$
F(z)\equiv \int_{-\infty}^z \frac{du}{(\sinh u/2)^{4/\kappa}}.
$$
The function $F(z)$ is well-defined and analytic on the strip
$\mathbb{S}_1$ for all $\kappa$'s. For $\kappa>4$, $F(z)$ is bounded
and has a continuous extension to the closure of $\mathbb{S}_1$. For
$\kappa<4$, it is unbounded near the origin and so are the
corresponding solutions to eq.(\ref{martineq}). In that case
these solutions only lead to local martingales and not true
martingales.  See Appendix for details on its definition and on its
properties.  We have $\Im{\rm m}\, F(\infty)=-\sin(2\pi/\kappa))\, I=
-\sin(4\pi/\kappa))\, J$ with
$$
I= \int_{-\infty}^{+\infty} dy
(\cosh y/2)^{-4/\kappa},\qquad
J= \int_0^\infty dy (\sinh y/2)^{-4/\kappa}.
$$
As a check one may verify that
$P_l(z,\bar z)$ behaves as expected on the boundary. On the positive
real axis, $(\sinh z/2)$ is real and positive so that
$$P_l(x)=0, \quad x\in \mathbb{R}_+,$$ 
in accordance with the fact that no point on the real axis can be on
the left of the trace. On the negative real axis, $(\sinh x/2)^{4/\kappa}=
e^{i4\pi/\kappa}(\sinh|x|/2)^{4/\kappa}$ and
$$ P_l(x)= 1 - \frac{1}{J}\int_{|x|}^{+\infty}
\frac{dy}{(\sinh y/2)^{4/\kappa}},\quad x\in \mathbb{R}_-,
$$
which interpolates between one and zero. It gives the probability of
the hull not to spread further than $x$ on the negative real axis.
On the upper boundary, 
$$ P_l(z=i\pi+x)= 1 - \frac{1}{I}\int_{-\infty}^x
\frac{dy}{(\cosh y/2)^{4/\kappa}}, \quad z\in i\pi+\mathbb{R},
$$
since there $(\sinh z/2)^{4/\kappa}=e^{2i\pi/\kappa}(\cosh
x/2)^{4/\kappa}$.
This yields the density probability for the trace to stops on an
interval $[x,x+dx]$ on the upper boundary.
\medskip

{\it --- Case (ii): probability to be inside the hull.}

The solution of the martingale equation (\ref{martineq}) satisfying the
boundary conditions (\ref{bdry2}) is given by the harmonic function
\begin{eqnarray}
P_{in}(z,\bar z)= \frac{\Im {\rm m}[e^{i2\pi/\kappa}\, F(z)\,]}{
\Im {\rm m}[e^{i2\pi/\kappa}\, F(0)\,]}
\label{probin}
\end{eqnarray}
with identical function $F(z)$ as above and $\Im {\rm
  m}[e^{i2\pi/\kappa}\, F(0)\,] = -\sin(2\pi/\kappa)\, J$.
Again, $P_{in}$ has the expected behavior on the boundary. 
Since $e^{i2\pi/\kappa}\, F(z)$ is real on the upper boundary, we
have
$$ P_{in}(z,\bar z)=0,\quad z\in i\pi +\mathbb{R}, $$
in agreement with the fact that no point on the upper boundary can be 
swallowed. $P_{in}$ is even on the real axis and
$$ P_{in}(x)=  \frac{1}{J}\int_{|x|}^{+\infty}
\frac{dy}{(\sinh y/2)^{4/\kappa}},\quad x\in \mathbb{R}.$$
This is of course complementary to $P_l(x)$ for $x$ negative.
\medskip

{\it --- CFT interpretation.}

The correlation function (\ref{btob}) has a natural interpretation in
the Coulomb gas representation of CFTs: the weight zero primary field
$\Phi_0$ is simply the integral of the screening current.  Recall that
CFT with $c=1-12\alpha_0^2<1$ may be represented in terms of a free
bosonic field $\varphi(z)$ with a background charge $2\alpha_0$, see
refs.\cite{nienhuis,FF,Difbook}. The conformal weight of a state of
coulomb charge $\alpha$, or $2\alpha_0-\alpha$, is
$h(\alpha)=\frac{1}{2}\alpha(\alpha-2\alpha_0)$.  The weight
$h_{r;s}={[(r\kappa-4s)^2-(\kappa-4)^2]}/{16\kappa}$ corresponds to
the charge $\alpha_{r,s}=\alpha_0 -\frac{r}{2}\alpha_+
-\frac{s}{2}\alpha_-$ with $\alpha_\pm$ the two screening charges. In
the present case $c=1-6{(\kappa-4)^2}/{4\kappa}$ and the
correspondence is $\alpha_-=-2\sqrt{2/\kappa}$,
$\alpha_+=\sqrt{\kappa/2}$ and $2\alpha_0=\alpha_++\alpha_-$.
In particular $\alpha_{1;2}=-\alpha_-/2$ and
$\alpha_{0;1/2}=\alpha_0\pm\alpha_-/4$. 
The screening charges are such that the currents
$Q_\pm(z)=\exp(i\alpha_\pm\varphi(z))$ have weight one.
The operator $\Phi_0$ in eq.(\ref{btob}) is a linear combination of
the primitive of the screening current $Q_-$ and the identity operator, ie:
$$
\Phi_0(z,\bar z) = {\rm const'.}\,{\bf 1}
+ \Re {\rm e}\, [\, {\rm const.}\, \int^z dw\, Q_-(w)\, ]. 
$$
Indeed this operator has conformal weight zero, satisfies the
appropriate fusion rules and fulfills the charge conservation
requirement which demands that the sum of the coulomb charges of the
operators involve in the correlation function minus the background
charge belongs to the lattice generated by the screening charges.

\section{The limiting case $\kappa=4$.}\label{sec:limcase}

For $\kappa=4$ the SLE trace is a simple curve so that no point are
swallowed and $P_{in}=0$ for all points.  This case is marginal in the
sense that the integral defining $F(z)$ is only logarithmically
divergent. By extension, we have:
\begin{eqnarray}
P_l(z,\bar z)= \frac{1}{\pi} \Im {\rm m}\, 
\left[\log(\tanh \frac{z}{4})\right].
\label{probk=4}
\end{eqnarray}
This satisfies the martingale equation (\ref{martineq}) for $\kappa=4$ 
and the appropriate boundary conditions: $P_l(x\in\mathbb{R}_+)=0$ and
$P_l(x\in\mathbb{R}_-)=1$. Contrary to the cases $\kappa<4$, it is
discontinuous at the origin.
On the upper boundary the distribution of the trace is given by:
$$ P_l(i\pi +x)= 1 -\frac{2}{\pi} \arctan(e^{x/2}),\quad
x\in\mathbb{R}.$$

For $\kappa=4$, the Virasoro central charge is $c=1$ and $h_{1;2}=1/4$
and $h_{0;1/2}=1/16$. The probability (\ref{probk=4}) possesses a nice free
field CFT interpretation.  Central charge $c=1$ corresponds to bosonic
free field.  Let us denote by $X$ this field.  $h_{1;2}=1/4$ is the
conformal weight of the boundary vertex operator $V_{1;2}=\cos
X/\sqrt{2}$ which may be thought of as the boundary condition changing
operator intertwining two boundary intervals on which two different
Dirichlet boundary conditions are imposed.  $h_{0;1/2}=1/16$ is the
dimension of the twist field $\sigma$ which is the boundary condition
changing operator intertwining between Dirichlet and Neumann boundary
conditions. Thus the probability $P_l(z,\bar z)$ is proportional to
$$
\vev{ X(z,\bar z)}_{D;D;N}= \vev{\sigma(x_+)\sigma(x_-)
  X(z,\bar z) V_{1;2}(x_0)}, $$
where 'D;D;N' refers to Dirichlet
boundary conditions on the lower boundary $[x_-,x_0]$ and $[x_0,x_+]$,
but with a discontinuity at $x_0$ (ie. two D-branes at finite
distance), and Neumann boundary condition on the upper boundary
$[x_-,x_+]$ (ie. a space filling brane).  The fact $P_l(z,\bar z)$
satisfies the Dirichlet boundary conditions on the lower boundary is
clear by construction but one may verify that it actually satisfies
the Neumann boundary condition on the upper boundary. The fact that it
is a harmonic function is then a consequence of the free field
equation of $X$.  \bigskip

\section{Boundary excursion probabilities (all $\kappa$'s).}
\label{sec:boundexc} 

The computations of previous sections yield the density probability of
the hitting point on the upper boundary for $\kappa>4$. We now would
like to show that this formula actually applies to any value of
$\kappa$. So we look for the probability $P_{up}(\hat x)$,
$\hat x\in i\pi+\mathbb{R}$, that the SLE trace hits the upper boundary at a
point $\hat u\equiv i\pi+u$, $u>x$, on the right of 
$\hat x\equiv i\pi+x$.  By definition, this probability satisfies:
$$ P_{up}(-\infty)=1,\quad P_{up}(+\infty)=0.$$
To change gear we shall do the computation using conformal field
theory techniques.
Using the CFT martingales (\ref{martin}), or equivalently (\ref{OvevGbis}),
we aim at proving that this probability is given by the correlation
function
$$ 
\vev{\psi_{0;1/2}(x_-)\psi_{0;1/2}(x_+)\psi_0(\hat x)
  \psi_{1;2}(x_0)},
$$
where $\psi_0(\hat x)$ is a weight zero boundary conformal field
inserted on the upper boundary. Recall that in the strip geometry
$x_\pm=\pm\infty$ and $x_0=0$. Due to CFT fusion rules, there are two
possible choices for $\psi_0$~: either it is the identity operator or
it is the intertwiner between the Virasoro modules of weights
$h_{1;2}=\frac{6-\kappa}{2\kappa}$ and
$h_{1;0}=\frac{\kappa-2}{2\kappa}$. The existence of two choices makes
possible to fulfill the above two boundary conditions. These boundary
condition code for the behavior of the functions as $\hat x$ approaches
$x_-$ or $x_+$. In CFT correlation functions this behavior is
governed by fusing $\psi_0$ with $\psi_{0;1/2}$. In non unitary CFT
there is in general not much constraint on fusions. However, if
the product of the operators $\psi_{0;1/2}\psi_0$ is acting on the
state $\ket{\omega}$ created by $\psi_{1;2}$, then the null vector
equation $(-2L_{-2}+\frac{\kappa}{2}L_{-1}^2)\ket{\omega}=0$ imposes
  constraints on the fusion. In the present case, taking into account that
  the out-state is created by $\psi_{1;2}$, we get:
$$
\psi_{0;1/2}(y)\psi_0(\hat x)\ket{\omega} \simeq_{\hat x\to y}
c\, [\psi_{0;1/2}(y)+\cdots ]\ket{\omega} 
+\tilde c\, (y-\hat x)^{2/\kappa}\, [\psi_{0;3/2}(y)+\cdots]\ket{\omega}
$$
where the dots refer to the descendant operators.  The two fusion
coefficients $c$ and $\tilde c$ depend on which operator $\psi_0$ we
choose. The noticeable fact is that this operator product expansion is
regular for all values of $\kappa$~: the coefficient in front of
$\psi_{0;1/2}$ is constant and that in front of the $\psi_{0;3/2}$
vanishes as $\hat x\to y$. This allows us to choose the operator $\psi_0$
such that its fusion with $\psi_{0;1/2}(x_+)$ at point $x_+$ vanishes
but that with $\psi_{0;1/2}(x_-)$ at point $x_-$ tends to a constant. 
With this choice the correlation function satisfies:
$$
\vev{\psi_{0;1/2}(x_-)\psi_{0;1/2}(x_+)\psi_0(\hat x)
  \psi_{1;2}(x_0)}\to\cases{ 1,& if $\hat x\to x_-$;\cr
                     0,& if $\hat x\to x_+$;\cr}
$$
By the martingale property (\ref{martin}) or
(\ref{OvevGbis}), this correlation function but
with $\hat x$ replaced by its image by $f_t$ , ie.
by $f_t(\hat x)$, is a martingale. It is such that, at large time,
$$
\lim_{t\to\infty}\vev{\psi_{0;1/2}(x_-)\psi_{0;1/2}(x_+)\psi_0(f_t(\hat
  x)) \psi_{1;2}(x_0)} = {\bf 1}_{\{ {\hat x}\ \mathrm{on~
    the~left~of~}\gamma_{\mathrm {SLE}} \} },
$$
so that it projects on those events for which $\hat x$ is on the left
of the SLE trace. Taking the expectation of this equation using
the martingale property implies our claim that
\begin{eqnarray}
P_{up}(\hat x)= \vev{\psi_{0;1/2}(x_-)\psi_{0;1/2}(x_+)\psi_0(\hat x)
  \psi_{1;2}(x_0)}.
\label{Pup}
\end{eqnarray}

This CFT correlation function can be explicitly computed as the null
vector relation $(-2L_{-2}+\frac{\kappa}{2}L_{-1}^2)\ket{\omega}=0$
implies that it satisfies a second order differential equation. In the 
strip geometry the later reads:
$$
\left( \frac{\kappa}{2}\partial_x^2 + \tanh
  (\frac{x}{2})\,\partial_x\right) P_{up}(\hat x)=0,
$$
with $\hat x= i\pi+x $. Its solution is:
\begin{eqnarray}
P_{up}(\hat x)=1-\frac{1}{I}
\int_{-\infty}^x\frac{dy}{(\cosh y/2)^{4/\kappa}}.
\label{Pup2}
\end{eqnarray}
For $\kappa>4$, it of course coincides with the formula we found in
the previous sections, but the derivation we presented here is valid
for any $\kappa$. Surprisingly, the formula (\ref{Pup2}) shows no
transition as a function of $\kappa$, neither at $\kappa=4$ nor at
$\kappa=8$. For $\kappa=6$, $(1-P_{up})$ gives Cardy's crossing formula
\cite{Cardy} for the existence of a cluster percolating from the
boundary interval $[x_-,x_0]$ to the opposite boundary interval
$[x,x_+]$ in critical percolation.

Why harmonic solutions to eq.(\ref{martineq}) have a probabilistic
relevance remains to be explained. It would have been legitimate to
expect a simple formula also for the probability for a bulk point --
and not only a boundary point -- to be on the left of the SLE trace.
Although this probability is again given by the bulk-to-boundary
correlation function (\ref{btob}), it is not harmonic for $\kappa<4$
and we do not have a closed formula for it in this regime, as we do
not have a clear understanding of this transition. Actually the same
remark applies to the probability, meaningful for $4<\kappa<8$, that a
bulk point be swallowed from the left by the SLE trace~: it is not
harmonic.

\section{Numerical simulations of Ising excursion probabilities.}
\label{sec:simul}

We have simulated the Ising ferromagnet on a square lattice strip of
size $L \times 3L$ at the critical temperature (namely $T_c/J =
2/\log(1+\sqrt 2)$, where $J$ is the ferromagnetic interaction). On
the lower boundary spins, we applied a $-J$ magnetic field on the left
half of the boundary and a $+J$ field on the right half of the
boundary (this is equivalent to adding a raw of frozen spins below the
lower raw of the strip). On the horizontal direction we applied {\em
antiperiodic} boundary conditions, so that the minus phase on the left
and the plus phase on the right would connect gracefully. This suppresses
most finite size effects due to the finite length of the strip. The
upper boundary was left free.

To relax this system we used a cluster algorithm originally developed
for spin glasses~\cite{Houdayer} (using 64 configurations at one
temperature). This choice of algorithm was made because the code was
available without much programming effort. More common techniques
could have been used and would have probably lead to better running
performance.

\begin{figure}[htb]
  \begin{center}
    \includegraphics[width=0.8\textwidth]{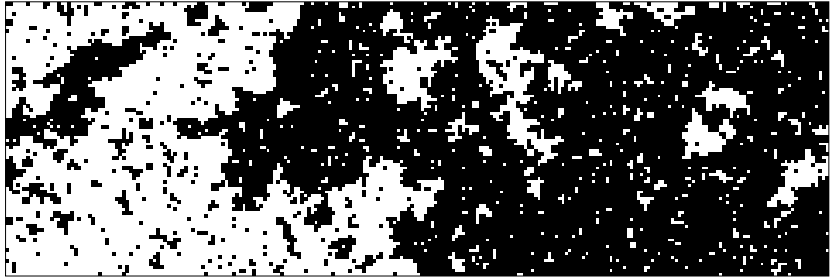}
      \caption{A sample equilibrium configuration at $L=80$, black
        and white squares respectively represent plus and minus spins.}
     \label{fig:config}
 \end{center}
\end{figure}

We have used different system sizes $L=10$, 14, 20, 26, 40 and 80 and
gathered for each size 320,000 independent samples (only 160,000 for
$L=80$). A sample configuration at $L=80$ is shown on
figure~\ref{fig:config}. For each sample, we measured the total lateral
displacement of the interface. More precisely we followed the interface
from the middle of the lower boundary up to the upper boundary and
counted how many times it went to the left or to the right (the
antiperiodic boundary may be crossed during this procedure). Note that
the interface is somewhat ill defined on the square lattice: it can
have branching points where it splits in two, in this case we chose
one of the branch with probability $1/2$. In some increasingly rare
cases (less than 0.03\% for the smaller size), it may also go around
the system (through the antiperiodic boundary) and comes back to its
starting point. In such cases we simply ignored the sample.

\begin{figure}[htbp]
  \begin{center}
    \includegraphics[width=0.8\textwidth]{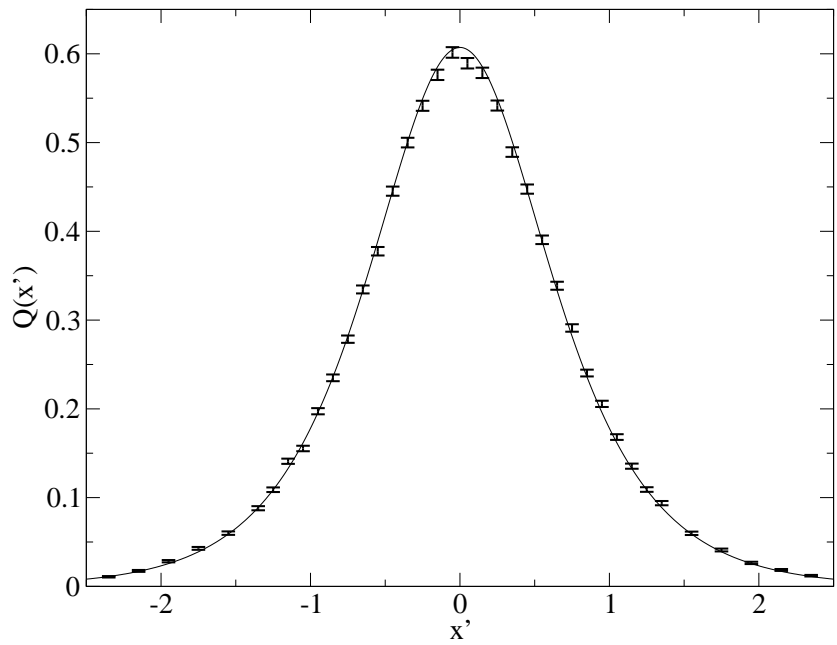}
      \caption{Reduced distribution of the Ising interface end point
        ($x'=x/L$). The plain curve is the analytical result, data
        points are numerical results for $L=80$.}
     \label{fig:distrib}
 \end{center}
\end{figure}

\begin{figure}[htbp]
  \begin{center}
    \includegraphics[width=0.8\textwidth]{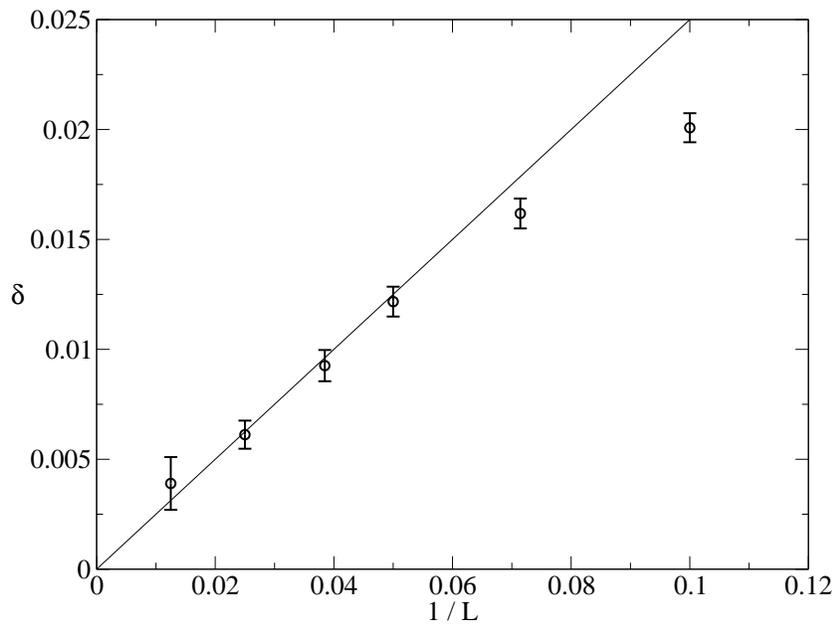}
      \caption{Value of the maximum absolute difference
        between the numerical and analytical integrated distributions
        of the Ising interface end point for $80\ge L\ge 10$.
        The straight line is just a guide to the eyes.}
     \label{fig:converge}
 \end{center}
\end{figure}

We thus get the distribution of end point position $x$. The
distribution $Q(x')$ of $x'=x/L$ is to be compared to the theoretical
one computed previously (\ref{Pup2}) (with $\kappa=3$) for a strip of
width $\pi$. In the simulation variables, it reads:
\begin{equation}
Q(x')=\frac{\pi}I \cosh(x' \pi/2)^{-4/3}.
\end{equation}
Where $I$ is in this case $6\;\Gamma(2/3)\;\Gamma(5/6)/\sqrt\pi\simeq
5.17422$. The theoretical and numerical distributions (for $L=80$) are
shown on figure~\ref{fig:distrib}. They agree very well. To make a
more precise statement, we considered the integrated distribution (for
which the statistical error bars are much smaller) and computed the
maximum $\delta$ of the absolute difference between the theoretical
and numerical integrated distributions. The results are shown on
figure~\ref{fig:converge}. This difference converges nicely to zero at
large $L$. Moreover the finite size corrections seems to be of order
$1/L$. Note that the finite size corrections were expected to be at a
least of order $1/L$ since the lattice spacing has a relative size of
order $1/L$.

\bigskip

\section{Conclusion.}

Besides chordal and radial SLEs, {\it dipolar SLEs} are the only
possible simple SLEs on simply connected planar domains which satisfy
a left-right reflection symmetry. As illustrated in previous sections,
they provide a generalization of chordal SLEs with more structures
since the hulls do not totally fill the domain for $\kappa>4$. The
fact that the bulk visiting probabilities we computed are harmonic
functions is clearly related to the discrete harmonic explorer. The
latter was defined in ref.\cite{SchraShief} to provide a discrete
analogue of chordal SLE at $\kappa=4$. It clearly can be generalized
to yield a discrete analogue of dipolar SLE at $\kappa=4$ by dealing
with harmonic functions with 'D;D;N' boundary conditions to specify
the probabilities of the excursion processes.  

The harmonicity property of the visiting probabilities for $\kappa>4$
was unexpected from a CFT point of view.  The origin of the
non-harmonicity transition -- ie. the fact the probabilities for a
point to be on the left (or on the right) of the curve stop to be
harmonic for $\kappa<4$ -- is not clear to us. We however noticed that
for $\kappa>4$ the probabilities for a point to be swallowed from the
left (or from the right) are also not harmonic functions. These two
breakdowns of harmonicity seem to be related by duality 
$\kappa\leftrightarrow 16/\kappa$.

Finally, numerical simulations of the Ising ferromagnet at criticality
confirms the analytical results presented here as well as the CFT mapping.

\vskip 1.5 truecm

{\bf Acknowledgements:} 
Work supported in part by EC contract number
HPRN-CT-2002-00325 of the EUCLID research training network.

\vskip 1.5 truecm

\appendix

\section{Appendix: analytical details on the function $F(z)$.}

Here we gathered a few informations on the function 
$$F(z)\equiv \int_{-\infty}^z \frac{du}{(\sinh u/2)^{4/\kappa}}.$$
We have to specify the analytical properties of $(\sinh
z/2)^{-4/\kappa}$. It is such that ${\rm arg}(\sinh z/2)\in[0,\pi]$.
Hence, $(\sinh z/2)^{-4/\kappa}$ is real on the positive real axis,
equals to $e^{-i4\pi/\kappa}(\sinh |x|/2)^{-4/\kappa}$ on the negative
real axis and equals to $e^{-i2\pi/\kappa}(\cosh x/2)^{-4/\kappa}$ on
the upper boundary $z=i\pi+x$. One may verify that it is equivalent to
$2^{4/\kappa}\,e^{-2z/\kappa}$ around $+\infty$, while it behaves as
$2^{4/\kappa}\,e^{(2z-i4\pi)/\kappa}$ around $-\infty$ .

By analyticity, the contour of integration in the definition of $F(z)$
can be choose arbitrary inside the strip $\mathbb{S}_1$ but starting
at $-\infty$. As a consequence, comparing the integration along the
real axis $\mathbb{R}$ and along $i\pi+\mathbb{R}$ we learn that
$F(0)= e^{-i4\pi/\kappa}\,J$ and that
$$ F(+\infty)=e^{-i2\pi/\kappa}\ I = e^{-i4\pi/\kappa}\ J + J,$$
where $I=\int_{-\infty}^{+\infty}dy(\cosh y/2)^{-4/\kappa}$ and 
$J=\int_{0}^{+\infty}dy(\sinh y/2)^{-4/\kappa}$ as in the text.
This leads to $I=2J\cos(2\pi/\kappa)$

For $z=a+ib$, $b\in[0,\pi]$, one may expand $\sinh z/2$ as
$$\sinh z/2= \sinh a/2\, \cos b/2 + i \cosh a/2
\sin b/2,$$ so that $(\sinh z/2)^{-4/\kappa}= R^{-4/\kappa}_z\,
\exp(-i4\theta_z/\kappa)$ with
\begin{eqnarray}
R_z &=& | \sinh z/2|, \nonumber\\
\tan \theta_z &=& \coth a/2\, \tan b/2,\quad \theta_z\in [0,\pi]. 
\nonumber
\end{eqnarray}
The angle $\theta_{a+ib}$ decreases from $(\pi-b/2)$ to $b/2$ as
$a$ varies from $-\infty$ to $+\infty$. 

The functions $\Im{\rm m}[F(z)]$ and $\Im{\rm
  m}[e^{i2\pi/\kappa}\,F(z)]$ involved in
eqs.(\ref{probleft},\ref{probin}), can be represented as:
\begin{eqnarray}
\Im{\rm m}[F(z=x+ib)]&=& - \int_{-\infty}^x da\, R^{-4/\kappa}_{a+ib}\,
\sin(4\theta_{a+ib}/\kappa), \nonumber\\
\Im{\rm m}[e^{i2\pi/\kappa}\,F(z=x+ib)]&=&
 -\int_{-\infty}^x da\, R^{-4/\kappa}_{a+ib}\,
\sin((4\theta_{a+ib}-2\pi)/\kappa). \nonumber
\end{eqnarray}
Other representations can be written using different integration contours.


\end{document}